\documentclass[prb,         showpacs,amsmath,amssymb,twocolumn]{revtex4}

\usepackage{dcolumn}
\usepackage{verbatim} 
\usepackage{graphicx}
\usepackage{epsfig}
\usepackage{hyperref} 

\begin{document}
\newcommand{\boldnabla}{\mbox{\boldmath$\nabla$}}

\title{Anderson localization as position-dependent diffusion in disordered waveguides}

\author{Ben Payne$^1$, Alexey Yamilov$^1$\footnote{Electronic~address:~yamilov@mst.edu} and Sergey E. Skipetrov$^2$\footnote{Electronic~address:~Sergey.Skipetrov@grenoble.cnrs.fr}}
\affiliation{$^1$Department of Physics, Missouri University of Science \& Technology, Rolla, MO 65409\\
$^2$Universit\'{e} Joseph Fourier, Laboratoire de Physique et Mod\'{e}lisation des Milieux Condens\'{e}s, CNRS, 25 rue des Martyrs, BP 166, 38042 Grenoble, France}

\date{\today}

\begin{abstract}
We show that the recently developed self-consistent theory of Anderson localization with a position-dependent diffusion coefficient is in quantitative agreement with the supersymmetry approach up to terms of the order of $1/g_0^2$ (with $g_0$ the dimensionless conductance in the absence of interference effects) and with  large-scale {\it ab-initio} simulations of the classical wave transport in disordered waveguides, at least for $g_0 \gtrsim 0.5$. In the latter case, agreement is found  even in the presence of absorption. Our numerical results confirm that in open disordered media,  the onset of  Anderson localization can be viewed as position-dependent diffusion.
\end{abstract}


\maketitle


\section{Introduction}
\label{sec:intro}

Anderson localization is a paradigm in condensed matter physics \cite{1958_Anderson}. It consists in a blockade of the diffusive electronic transport in disordered metals due to interferences of multiply scattered de Broglie waves at low temperatures and at a sufficiently strong disorder. This phenomenon is not unique to electrons but can manifest itself for any wave in the presence of disorder, in particular for classical waves, such as light and sound \cite{1984_John_prl}, and, as shown more recently, for matter waves \cite{billy08}. Although the absence of decoherence and interactions \cite{2007_Akkermans_book} for classical waves is appealing in the context of the original idea of Anderson, serious complications appear due to absorption of a part of the wave energy by the disordered medium \cite{1991_Genack}.
Extracting clear signatures of Anderson localization from experimental signals that are strongly affected by --- often a poorly controlled --- absorption was the key to success in recent experiments with microwaves \cite{2000_chabanov_nature,2003_Genack}, light \cite{2006_Maret_PRL} and ultrasound \cite{2008_van_Tiggelen_Nature}.

Classical waves offer a unique possibility of performing angle-, space-, time- or frequency-resolved measurements with excellent resolution, the possibility that was not available in the realm of electronic transport. In a wider perspective, they also allow a controlled study of the interplay between disorder and interactions, as illustrated by the recent work on disordered photonic lattices \cite{schwartz07}.
Interpretation of measurements requires a theory that would be able to describe not only the genuine interferences taking place in the bulk of a large sample but also the modification of these interferences in a sample of particular shape, of finite size, and with some precise conditions at the boundaries. Such a theory has been recently developed \cite{2000_van_Tiggelen,2004_Skipetrov,2006_Skipetrov_dynamics,2008_Cherroret} based on the self-consistent (SC) theory of Vollhardt and W\"{o}lfle \cite{1980_Vollhardt_Wolfle}. The new ingredient is the position dependence of the renormalized diffusion coefficient $D(\mathbf{r})$ that accounts for a stronger impact of interference effects in the bulk of the disordered sample as compared to the regions adjacent to boundaries. This position dependence is crucial in open disordered media \cite{2009_Cherroret}. $D(\mathbf{r})$ also appears in the supersymmetry approach to wave transport \cite{2008_Tian}, which confirms that this concept goes beyond a particular technique (diagrammatic or supersymmetry methods) used in the calculations.

The SC theory with a position-dependent diffusion coefficient was successfully applied to analyze microwave \cite{2004_Skipetrov} and ultrasonic \cite{2008_van_Tiggelen_Nature} experiments. The predictions of the theory \cite{2006_Skipetrov_dynamics} are also in qualitative agreement with optical experiments of St\"{o}rzer \textit{et al.} \cite{2006_Maret_PRL}. However, it remains unclear whether the position dependence of $D$ is just a (useful) mathematical concept or if it is a genuine physical reality. In addition, the extent to which predictions of SC theory are quantitatively correct is not known. Obviously, the last issue is particularly important once comparison with experiments is attempted.

In the present paper we compare the predictions of SC theory of localization with the known results obtained previously using the supersymmetry method \cite{2000_Mirlin} and with the results of extensive \textit{ab-initio} numerical simulations of wave transport in two-dimensional (2D) disordered waveguides. We demonstrate, first, that the position-dependent diffusion is a physical reality and, second, that SC theory agrees with the supersymmetry approach up to terms of the order of $1/g_0^2$ (with with $g_0$ the dimensionless conductance in the absence of interference effects) and with numerical simulation at least for $g_0 \gtrsim 0.5$. In the latter case, the agreement is found even in the presence of absorption.

\section{Self-consistent theory of localization}
\label{sec:sctheory}

We consider a scalar, monochromatic wave $u(\mathbf{r})e^{-i \omega t}$ propagating in a 2D volume-disordered waveguide of width $w$ and length $L \gg w$. The wave field $u(\mathbf{r})$ obeys the 2D Helmholtz equation:
\begin{equation}
\left\{\nabla^2 + k^2\left[1 + i \epsilon_a + \delta\epsilon(\mathbf{r}) \right]\right\} u(\mathbf{r}) = 0.
\label{eq:helmholtz}
\end{equation}
Here $k=\omega/c$ is the wavenumber, $c$ is the speed of the wave in the free space, $\epsilon_a$ is the imaginary part of the dielectric constant accounting for the (spatially uniform) absorption in the medium, and $\delta\epsilon(\mathbf{r})$ is the randomly fluctuating part of the dielectric constant.
Assuming that $\delta\epsilon(\mathbf{r})$ is a Gaussian random field with a short correlation length, it is easy to show that the disorder-averaged Green's function of Eq.\ (\ref{eq:helmholtz}), $\langle G(\mathbf{r}, \mathbf{r}') \rangle$, decays exponentially with the distance $|\mathbf{r}-\mathbf{r}'|$
\cite{2007_Akkermans_book}. The characteristic length of this decay defines the mean free path $\ell$. In this paper we consider quasi-1D waveguides defined by the condition $w \lesssim \ell \ll L$. The intensity Green's function of Eq.\ (\ref{eq:helmholtz}), $C(\mathbf{r}, \mathbf{r}') = (4\pi/c)
\langle \left| G(\mathbf{r}, \mathbf{r}') \right|^2 \rangle$, obeys self-consistent equations that can be derived following the approach of Ref.~\citenum{2008_Cherroret}. In a quasi-1D waveguide, all position-dependent quantities become functions of the longitudinal coordinate $z$ only and the stationary SC equations can be written in a dimensionless form:
\begin{eqnarray}
&&\left[\beta^2 - \frac{\partial}{\partial \zeta} d(\zeta)
 \frac{\partial}{\partial \zeta} \right] {\hat C}(\zeta,\zeta')
= \delta(\zeta-\zeta'),
\label{eq:sceq1}
\\
&&\frac{1}{d(\zeta)} =  1+\frac{2}{{\tilde g}_0}
{\hat C}(\zeta,\zeta).
\label{eq:sceq2}
\end{eqnarray}
Here ${\hat C}(\zeta,\zeta') = (w D_0/L)C(\mathbf{r},\mathbf{r}')$,
$D_0 = c\ell/2$ is the Boltzmann diffusion coefficient, $\zeta = z/L$ is the dimensionless coordinate, $d(\zeta) = D(z)/D_0$ is the normalized position-dependent diffusion coefficient, $\beta = L/L_a$ is the absorption coefficient (with $L_a = \sqrt{\ell \ell_a/2}$ and $\ell_a = 1/k\epsilon_a$ the macro- and microscopic absorption lengths, respectively), and ${\tilde g}_0 = (\pi/2)N \ell/L$ with $N = kw/\pi$ the number of the transverse modes in the waveguide. These equations should be solved with the following boundary conditions:
\begin{eqnarray}
{\hat C}(\zeta,\zeta^{\prime}) \mp
\frac{z_0}{L} d(\zeta) \frac{\partial}{\partial \zeta}
{\hat C}(\zeta,\zeta^{\prime}) = 0
\label{eq:bc}
\end{eqnarray}
at $\zeta = 0$ and $\zeta = 1$. Similarly to the 3D case \cite{2008_Cherroret}, these conditions follow from the requirement of vanishing incoming diffuse flux at the open boundaries of the sample. $z_0$ is the so-called extrapolation length equal to $(\pi/4)\ell$ in the absence of internal reflections at the sample surfaces \cite{1999_van_Rossum}. We will use $z_0 = (\pi/4) \ell$ throughout this paper. When Eqs.\ (\ref{eq:sceq1}--\ref{eq:bc}) are solved in the diffuse regime ${\tilde g}_0 \gg 1$, the dimensionless conductance of the waveguide is found to be $g_0 = (\pi/2)N \ell/(L + 2 z_0)$  \cite{1999_van_Rossum,1997_Beenakker} which is close to ${\tilde g}_0$ for $z_0 \ll L$.

In the absence of absorption ($\beta = 0$) we can simplify Eq.\ (\ref{eq:sceq1}) by introducing $\tau = F(\zeta) = \int_0^{\zeta} d\zeta_1/d(\zeta_1)$:
\begin{eqnarray}
-\frac{\partial^2}{\partial \tau^2} {\hat C}(\tau, \tau^{\prime})
= \delta(\tau-\tau^{\prime}),
\label{eq:sceq3}
\end{eqnarray}
with the boundary conditions (\ref{eq:bc}) becoming
\begin{eqnarray}
{\hat C}(\tau, \tau^{\prime}) \mp
\tau_0 \frac{\partial}{\partial \tau}
{\hat C}(\tau, \tau^{\prime}) = 0,
\label{eq:bc3}
\end{eqnarray}
and $\tau^{\prime} = F(\zeta^{\prime})$, $\tau_0 = z_0/L$. Equations (\ref{eq:sceq3}) and (\ref{eq:bc3}) are readily solved:
\begin{eqnarray}
{\hat C}(\tau, \tau^{\prime}) =
\frac{(\tau_< + \tau_0)(\tau_{\mathrm{max}} + \tau_0 - \tau_>)}{\tau_{\mathrm{max}} + 2 \tau_0},
\label{eq:sol}
\end{eqnarray}
where $\tau_< = \min(\tau, \tau^{\prime})$, $\tau_> = \max(\tau, \tau^{\prime})$ and $\tau_{\mathrm{max}} = F(1)$.  We now substitute this solution into Eq.\ (\ref{eq:sceq2}) to obtain
\begin{eqnarray}
\frac{1}{d(\tau)} \equiv \frac{d \tau}{d\zeta} =
1 + \frac{2}{\tilde g}_0 \times \frac{(\tau + \tau_0)(\tau_{\mathrm{max}} + \tau_0 - \tau)}{\tau_{\mathrm{max}} + 2 \tau_0}.
\label{eq:dtdz}
\end{eqnarray}
This differential equation can be integrated to find $\tau$ as a function of $\zeta$. Using $d(\zeta) = (d\tau/d\zeta)^{-1}$ we finally find
\begin{eqnarray}
d(\zeta) &=& \left\{ {\tilde g}_0 \sqrt{p} \cosh(\sqrt{p} \zeta/{\tilde g}_0) \right.
\nonumber \\
&-& \left. [{\tilde g}_0 + \tau_0(1 - p)] \sinh(\sqrt{p} \zeta/{\tilde g}_0) \right\}^2
\nonumber \\
&\times& \left\{ p [({\tilde g}_0 + \tau_0)^2 - \tau_0^2 p]
\right\}^{-1},
\label{eq:dzsol}
\end{eqnarray}
where $p$ is the solution of a transcendental equation
\begin{eqnarray}
\frac{2 {\tilde g}_0}{\sqrt{p}}
\mathrm{arctanh} \left\{ \frac{1}{\sqrt{p}}
\left[ 1 - \frac{\tau_0}{{\tilde g}_0} \left(p - 1 \right) \right] \right\} = 1.
\label{eq:p}
\end{eqnarray}
Solving the last equation numerically and substituting the result into Eq.\ (\ref{eq:dzsol}) we can find the profile $d(\zeta)$ at any ${\tilde g}_0$ and $\tau_0 = z_0/L$. In contrast,
for $\beta > 0$ Eqs.\ (\ref{eq:sceq1}--\ref{eq:bc}) do not admit analytic solution and we solve them by iteration: we start with $D(z) = D_0$, solve Eq.\ (\ref{eq:sceq1}) numerically with the boundary conditions (\ref{eq:bc}) and then find the new $D(z)$ from Eq.\ (\ref{eq:sceq2}). This procedure is then repeated until it converges to a solution. In typical cases considered in this paper the convergence is achieved after 10--20 iterations.

The simplest object that Eqs.\ (\ref{eq:sol}--\ref{eq:dzsol}) allows us to study is the average conductance of the waveguide $\langle g \rangle$. Indeed, the average transmission coefficient of the waveguide is found as
\begin{eqnarray}
T &=& -D(L) \left. \frac{dC(z, z^{\prime}=\ell)}{dz} \right|_{z=L}
\nonumber \\
&=& - \frac{1}{w} \times \left.
\frac{d{\hat C}(\tau, \tau_{\ell})}{d\tau}
\right|_{\tau = \tau_{\mathrm{max}}}
\nonumber \\
&=& \frac{1}{w} \times \frac{\tau_{\ell} + \tau_0}{\tau_{\mathrm{max}} + 2 \tau_0},
\label{eq:t}
\end{eqnarray}
where $\tau_{\ell} = F(\ell/L)$.
For the waveguide we have $\langle g \rangle \propto T$. A ratio that emphasizes the impact of localization effects is $\langle g \rangle/g_0 = T/T_0$, where $T_0$ is the average transmission coefficient found in the absence of localization effects (i.e., for $d \equiv 1$): $T_0 = (\ell + z_0)/w(L + 2 z_0)$. We find
\begin{eqnarray}
\frac{\langle g \rangle}{g_0} =
\frac{L + 2 z_0}{\ell + z_0}
(\tau_{\ell} + \tau_0)
\frac{p - 1}{2 {\tilde g}_0}.
\label{eq:goverg0}
\end{eqnarray}

Simple analytic results follow for $z_0 = 0$, when $g_0 = {\tilde g}_0$. Equation (\ref{eq:dzsol}) yields
\begin{eqnarray}
d(\zeta) &=& \left[ \frac{\sinh(\sqrt{p} \zeta/g_0)}{\sqrt{p}} - \cosh(\sqrt{p} \zeta/g_0) \right]^2
\label{eq:dzsol2}
\end{eqnarray}
and we find
\begin{eqnarray}
\tau_{\ell} &=& \frac{g_0}{\sqrt{p}\; \mathrm{cotanh}
(\sqrt{p} \ell/L g_0) -1}.
\label{eq:tauell}
\end{eqnarray}
In the weak localization regime $g_0 \gg 1$ the solution $p$ of Eq.\ (\ref{eq:p}) can be found as a series expansion in powers of $1/g_0$:
$p = 2 g_0 + 1/3 + 2/45 g_0 - 17/540 g_0^2 + \ldots$.
If we keep only the first term $p = 2 g_0$, substitute it into Eq.\ (\ref{eq:dzsol2}) and expand in powers of $1/g_0 \ll 1$, we obtain
$D(z) \simeq D_0 [ 1 - (2/g_0) (z/L)(1-z/L)]$.
Keeping terms up to $1/g_0^2$ in the expression for $p$ and substituting it into Eqs.\ (\ref{eq:tauell}) and (\ref{eq:goverg0}), expanding the result in powers of $1/g_0$ and then taking the limit of $L/\ell \rightarrow \infty$, we obtain
\begin{eqnarray}
\frac{\langle g \rangle}{g_0} \simeq
1 - \frac{1}{3 g_0} + \frac{1}{45 g_0^2} + \frac{2}{945 g_0^3} + \ldots.
\label{eq:goverg01}
\end{eqnarray}
This result coincides {\em exactly} with Eq.\ (6.26) of Ref.\ \onlinecite{2000_Mirlin} obtained by Mirlin using supersymmetry approach, except for a factor of 2 due to two independent spin states of electrons in Ref.\ \onlinecite{2000_Mirlin}. We therefore proved the exact equivalence between SC theory and the supersymmetry approach for the calculation of the average conductance $\langle g \rangle$ up to terms of the order of $1/g_0^2$.

Deep in the localized regime $g_0 \ll 1$ and Eq.\ (\ref{eq:p}) can be solved approximately to yield
$p = 1 + 4 \exp(-1/g_0)$ (always for $z_0 = 0$ and hence for $g_0 = {\tilde g}_0$).
If we substitute this $p$ into Eq.\ (\ref{eq:dzsol2}), we obtain
$D(z) \simeq D_0 \{ \exp(-z/\xi) + \exp[-(L-z)/\xi] \}^2$, where $\xi = g_0 L$ is the localization length.
Equations (\ref{eq:tauell}) and (\ref{eq:goverg0}) then yield
\begin{eqnarray}
\frac{\langle g \rangle}{g_0} \simeq
\frac{2}{g_0} \exp\left(-\frac{1}{g_0} \right),
\label{eq:goverg02}
\end{eqnarray}
where we made use of the fact that $L/\ell \gg 1$ and $N \gg 1$.
In contrast to Eq.\ (\ref{eq:goverg01}), this result differs from the one obtained using the supersymmetry approach [see Eq.\ (6.29) of Ref.\ \onlinecite{2000_Mirlin}]. Even though the exponential decay of conductance with $1/g_0 = L/\xi$ --- expected in the localized regime --- is reproduced correctly, both the rate of this decay and the pre-exponential factor are different. We thus conclude that SC theory does not provide quantitatively correct description of stationary wave transport in disordered waveguides in the localized regime.

It is worthwhile to note that the breakdown of SC theory for $g_0 \ll 1$ is not surprising and could be expected from previous results. Indeed, it has already been noted that for the time-dependent transmission, SC theory does not apply after the Heisenberg time $t_H$ \cite{2004_Skipetrov}. The stationary transmission coefficient $T$ of Eq.\ (\ref{eq:t}) is an integral of the time-dependent transmission $T(t)$: $T = \int_0^{\infty} dt\; T(t)$, with the peak of $T(t)$ around the Thouless time $t_D = L^2/\pi^2 D_0$ \cite{2004_Skipetrov}. When $g_0 \sim t_H/t_D \gg 1$, the integral is dominated by $t < t_H$ where SC theory applies. The integration thus yields the correct $T$. However, when $g_0 \ll 1$, $t_H$ is smaller than $t_D$ and the main part of pulse energy arrives at $t > t_H$. Such long times are beyond the reach of SC theory, hence its breakdown for small $g_0$.

\section{Numerical model}
\label{sec:numerical}

To test the predictions of the SC model discussed in the previous section we solve Eq.~(\ref{eq:helmholtz}) numerically using the method of transfer matrices defined in the basis of the transverse modes of the empty waveguide \cite{2007_Froufe-Perez_PRE,2010_Payne_closed}. To this end, we represent $\delta \epsilon(\mathbf{r})$ as a collection of $M$ randomly positioned ``screens'' perpendicular to the axis $z$ of the waveguide and characterized by random functions $f_{\nu}(y) = \sum_{n=1}^N \chi_n(y)\chi_n(y_\nu)$:
\begin{equation}
\delta\epsilon(\mathbf{r}) = \alpha\sum\limits_{\nu=1}^M \delta(z - z_\nu)f_\nu(y).
\label{de}
\end{equation}
Here $\chi_n(y) = (2/w)^{1/2}\sin(\pi ny/w)$ are the transverse modes of the waveguide and $y_\nu$ are chosen at random within the interval $(0, w)$. $z_\nu$ represent random positions of the screens, whereas $\alpha$ measures their scattering strength. Absorption can be included in the model by making $\alpha$ complex.

\begin{figure*}
\vskip -0.3in
\centering{\includegraphics[height=6.9in,angle=-90]{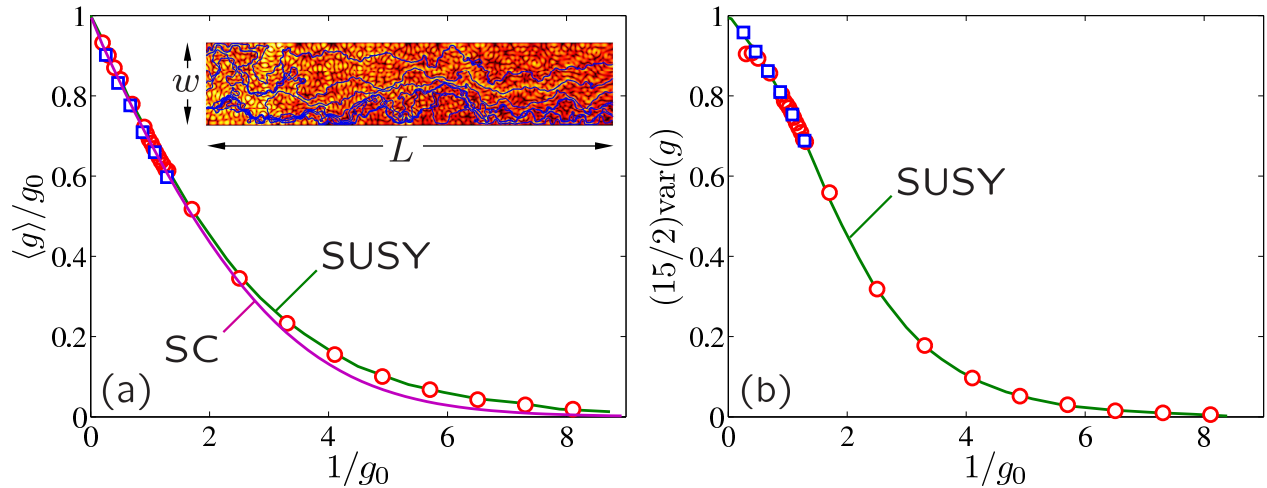}}
\vskip -2.2in
\caption{\label{fig1}
The average (a) and the variance (b) of the conductance $g$ of disordered waveguides supporting $N = 10$ (circles) and $N = 20$ (squares) modes are shown versus the inverse of $g_0$. The solid lines marked as SUSY are fits using Eq.\ (6.23) of Ref.\ \onlinecite{2000_Mirlin}, derived using the supersymmetry approach, with $\ell = 15.7 \lambda$ as the only fit parameter. The solid line marked as SC in (a) is obtained using the self-consistent theory [Eq.\ (\ref{eq:goverg0})]. Inset in (a): for a given realization of disorder, wave ``trajectories'' found by connecting local Poynting vectors are superimposed on the distribution of intensity $|u(\mathbf{r})|^2$ in a disordered waveguide with $w=10.25\lambda$ and $L=50\lambda$. Only trajectories that traverse the waveguide are shown.}
\end{figure*}

In the limit $N\rightarrow\infty$, $f_\nu(y)$ becomes a delta-function $\delta\left(y-y_\nu\right)$, mimicking a point-like scatterer. By the choice of $f_\nu(y)$ in Eq.\ (\ref{de}) we narrowed the basis to $N$ right- and $N$ left-propagating modes with real values of the longitudinal component of the wavevector. Such modes are often termed ``open channels'' in the literature \cite{2007_Froufe-Perez_PRE}. Hence, the total transfer matrix of the system is a product of $M$ pairs of $2N\times 2N$ scattering matrices corresponding to the random screens positioned at $z_{\nu}$ and the free space in between them, respectively \cite{2010_Payne_closed}. Because the numerical computation of products of a large number of transfer matrices ($\sim 10^2$--$10^5$ for the results in this paper) is intrinsically unstable, we implement a self-embedding procedure \cite{1999_yamilov_selfembed} which limits the errors in flux conservation to less than $10^{-10}$ in all cases. The system is excited by illuminating the waveguide with $N$ unit fluxes (one in each right propagating mode) and the wave field $u(\mathbf{r})$ is computed \cite{1999_yamilov_selfembed,2010_Payne_closed} for a given realization of disorder [see the inset of Fig.~\ref{fig1}(a)]. To compute statistical averages, ensembles of no fewer than $10^7$ realizations are used.

To estimate the mean free path $\ell$ of waves in our model system we perform a set of simulations for different disorder strengths and waveguide lengths, exploring both the regime of classical diffusion ($g_0 > 1$) and that of Anderson localization ($g_0 < 1$). The results of the simulations are used to compute the dimensionless conductance $g$, equal to the sum of all outgoing fluxes at the right end of the waveguide, and then to study its average value $\langle g \rangle$ and variance $\mathrm{var}(g)$ \cite{2006_Yamilov_conductance}. The dependencies of $\langle g \rangle$ and $\mathrm{var}(g)$ on $g_0$ are fitted by the analytic expressions obtained by Mirlin \cite{2000_Mirlin} using the supersymmetry approach, with $\ell$ as the only fit parameter (Fig.\ \ref{fig1}) \cite{2010_Payne_closed}. The best fit is obtained with $\ell = (15.7\pm 0.2) \lambda$.
In Fig.\ \ref{fig1}(a) we also show Eq.\ (\ref{eq:goverg0}) following from SC theory. As could be expected from the discussion in the previous section, the prediction of SC theory coincides with both the results of the supersymmetry approach and numerical simulations only for large $g_0 \gtrsim 0.5$.

\section{Position-dependent diffusion coefficient}
\label{sec:position}

\begin{figure}
\vskip -0.5cm
\centering{\includegraphics[height=3.5in,angle=-90]{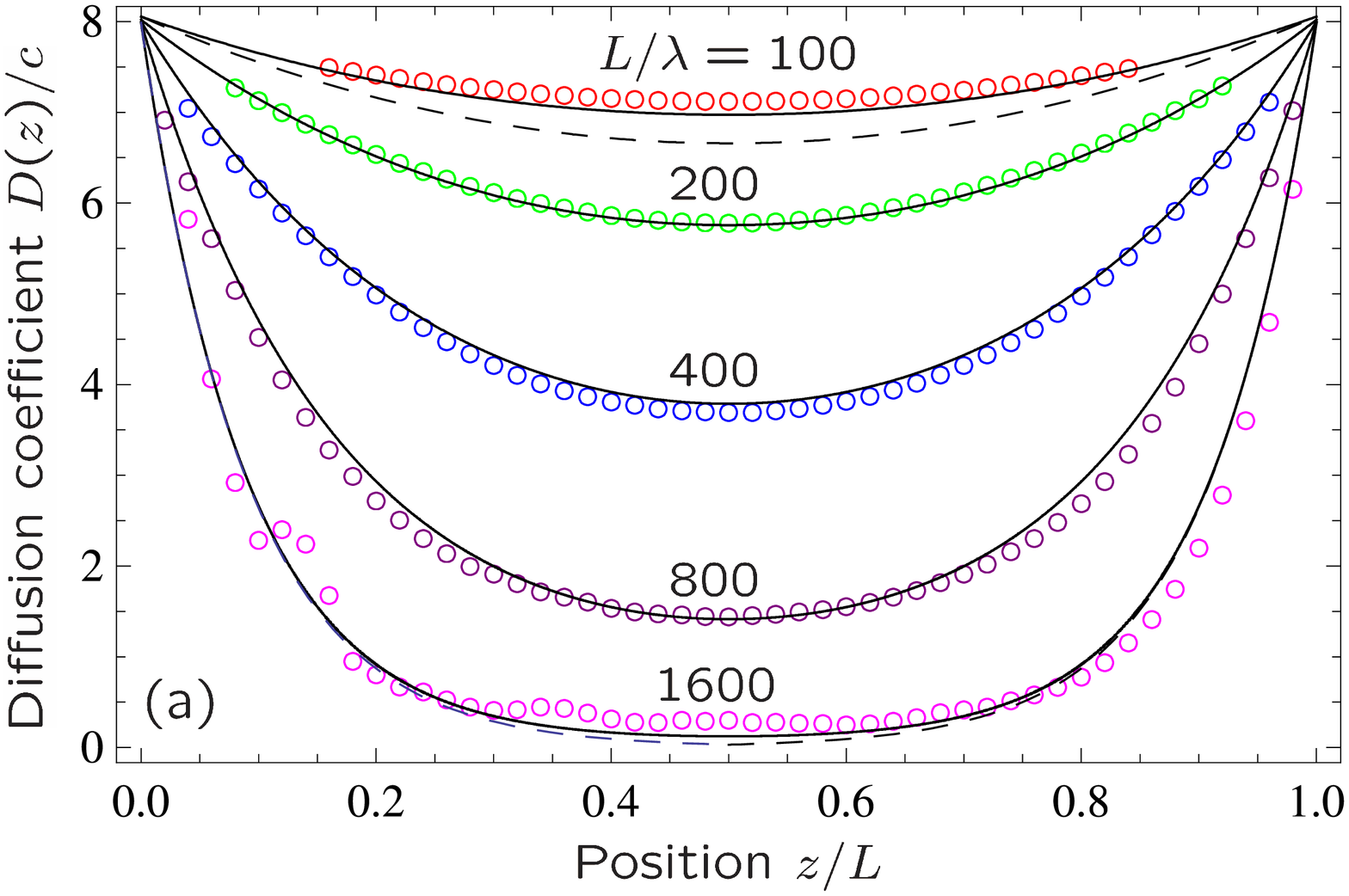}
\vskip -0.7cm
           \includegraphics[height=3.5in,angle=-90]{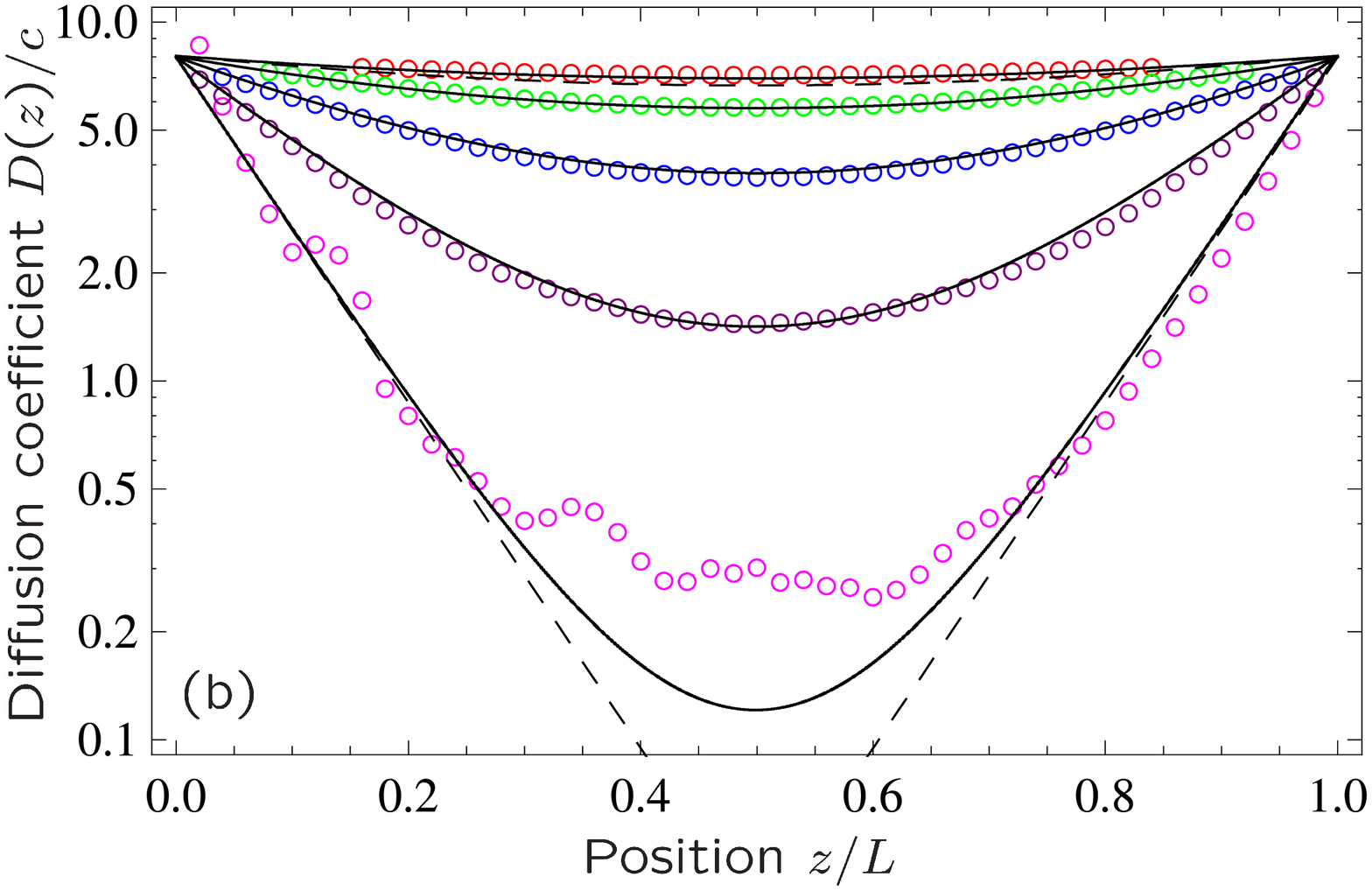}}
\vskip -0.7cm
\caption{\label{fig2} (a) Position-dependent diffusion coefficient $D(z)$ in 2D waveguides supporting the same number $N = 10$ of transverse modes (width $w=5.25\lambda$) but having different lengths $L$. Disorder is the same for all lengths. Symbols show the results of numerical simulations, whereas solid lines are obtained from the self-consistent theory with the mean free path $\ell = 17.5\lambda$. Dashed lines show the approximate results for
$g_0 \gg 1$ (shown for $L = 100 \lambda$) and $g_0 \ll 1$ (shown for $L = 1600 \lambda$), with $D(0)$ substituted for $D_0$, see text. (b) Same as (a) but in the logarithmic scale.}
\end{figure}

The wave field $u(\mathbf{r})$ that we obtain as an outcome of the numerical algorithm allows us to calculate the energy density ${\cal W}(\mathbf{r})$ and flux $\mathbf{J}(\mathbf{r})$ \cite{1953_Morse}:
\begin{eqnarray}
{\cal W}({\bf r}) &=& \frac{k^2}{2}\left|u({\bf r})\right|^2 + \frac{1}{2}\left| \boldnabla u({\bf r})\right|^2, \label{eq:W}\\
{\bf  J}({\bf r}) &=& -kc\; \mathrm{Im} \left[u({\bf r})\boldnabla u({\bf r})\right]. \label{eq:J}
\end{eqnarray}
These two quantities formally define the diffusion coefficient $D(z)$ which, in general, may be position-dependent:
\begin{equation}
D(z) = -\frac{\langle J_z(\mathbf{r}) \rangle}{\frac{d}{d z} \langle{\cal W}(\mathbf{r}) \rangle},
\label{eq:Dofz_definition}
\end{equation}
where the averages $\langle \ldots \rangle$ are taken over a statistical ensemble of disorder realizations as well as over the crossection of the waveguide. Eq.\ (\ref{eq:Dofz_definition}) can be used only at distances beyond one mean free path $\ell$ from the boundaries of the random medium because more subtle propagation effects of non-diffusive nature start to be important in the immediate vicinity of the boundaries \cite{2007_Akkermans_book}.

We first consider non-absorbing disordered waveguides described by $\epsilon_a = 0$ in Eq.\ (\ref{eq:helmholtz}) and real $\alpha$ in Eq.\ (\ref{de}).
In Fig.\ \ref{fig2} we compare numerical results for $D(z)$ with the outcome of SC theory for waveguides of different lengths but with statistically equivalent disorder. Quantitative agreement is observed for $L = 100$--$800 \lambda$, corresponding $g_0 \approx 0.3$--2. For the longest of our waveguides ($L = 1600\lambda$, $g_0 \approx 0.16$), deviations of numerical results from SC theory start to become visible in the middle of the waveguide, which is particularly apparent in the logarithmic plot of Fig.\ \ref{fig2}(b).
The mean free path $\ell = 17.5\lambda$ corresponding to the best fit of SC theory to numerical results is only about $10\%$ higher than $\ell = 15.7 \lambda$ obtained from the fits in Fig.\ \ref{fig1}.

We checked that the results of numerical simulations are not sensitive to the microscopic details of disorder: $D(z)$ obtained in two runs with different scattering strengths $\alpha$ and different scatterer densities, but equal mean free paths $\ell$ turned out to be the same.

\section{Effect of absorption}
\label{sec:absorption}

The linear absorption is modeled by introducing a non-zero $\epsilon_a$ in Eq.\ (\ref{eq:helmholtz}) and making $\alpha$ in Eq.\ (\ref{de}) complex. A link between $\epsilon_a$ and $\alpha$ can be established using the condition of flux continuity. Indeed, for continuous waves considered in this work the continuity of the flux leads to
\begin{equation}
\left\langle \boldnabla \cdot {\bf J}({\bf r})\right\rangle = (c/\ell_a) \left\langle{\cal W}({\bf r})\right\rangle,
\label{eq:flux_concervation}
\end{equation}
where $\ell_a = 1/k\epsilon_a$.
We checked that within numerical accuracy of our simulations the proportionality factor $c/\ell_a$ indeed remains constant independent of $z$. Therefore, Eq.\ (\ref{eq:flux_concervation}) allows us to determine the microscopic absorption length $\ell_a$ as
$c \langle{\cal W}({\bf r})\rangle/\langle \boldnabla \cdot {\bf J}({\bf r})\rangle$ obtained numerically at a given $\alpha$.

\begin{figure}
\vskip -0.5cm
\centering{\includegraphics[height=3.5in,angle=-90]{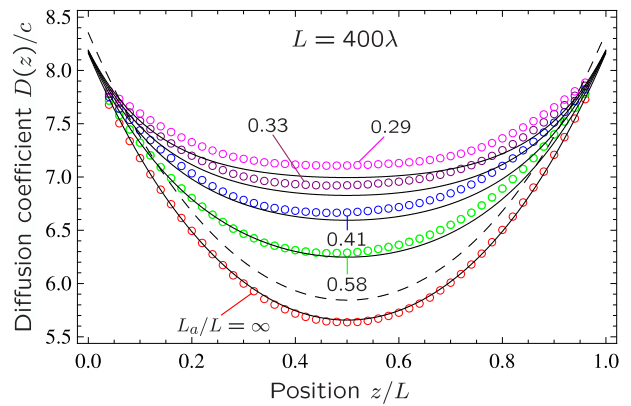}}
\vskip -0.7cm
\caption{\label{fig3} The effect of absorption on the position-dependent diffusion coefficient. Symbols are results of numerical simulations in a 2D waveguide of length $L = 400\lambda$, width $w = 10.25 \lambda$ ($N = 20$) and several values of the macroscopic absorption length $L_a$ indicated on the figure. Lines are obtained from SC theory with $\ell = 17.1 \lambda$ adjusted to obtain the best fit for the case of no absorption (lower curve). Dashed line shows $D(z)$ following from the self-consistent theory with the same $\ell = 17.5 \lambda$ as in Fig.~\ref{fig2} and illustrates the sensitivity of $D(z)$ to the exact value of $\ell$.}
\end{figure}

Figure\ \ref{fig3} demonstrates the effect of absorption on the position-dependent diffusion coefficient for a waveguide of length $L = 400\lambda$, which is about 25 mean free paths. For this waveguide $g_0 \simeq 1.3$ and the localization corrections are important. We observe that  absorption suppresses the localization correction to the position-dependent diffusion coefficient. This clearly demonstrates that the absorption nontrivially affects the transport by changing the way the waves interfere.
Nevertheless, we observe good agreement between numerical results (symbols) and SC theory (solid lines). The predictions of SC theory start to deviate from numerical results only for strong absorption ($L_a/L \lesssim 0.4$). Once again, the mean free path $\ell = 17.1 \lambda$ obtained from the fit of SC theory to the lower curve of Fig.\ \ref{fig3} is within 10\% of the value estimated from the variance of dimensionless conductance.

\section{Conclusions}
\label{sec:conclusions}

Two important results were obtained in this work. First, we convincingly demonstrated that the position-dependent diffusion coefficient is not an abstract mathematical concept but is a physical reality. The results of numerical simulations of scalar wave transport in disordered 2D waveguides unambiguously show that the onset of Anderson localization manifests itself as position-dependent diffusion. The reduction of the diffusion coefficient $D(\mathbf{r})$ is much more important in the middle of an open sample than close to its boundaries, in agreement with predictions of the self-consistent theory of localization.
Second, we established that for monochromatic waves in 2D disordered waveguides predictions of the self-consistent theory of localization are {\it quantitatively\/} correct provided that the dimensionless conductance in the absence of interference effects $g_0$ is at least larger than $0.5$. Moreover, the self-consistent theory yields a series expansion of the average conductance $\langle g \rangle$ in powers of $1/g_0$ that coincides exactly with the expansion obtained using the supersymmetry method \cite{2000_Mirlin} up to terms of the order of $1/g_0^2$. This was not obvious {\it a priori\/} because of the numerous approximations involved in the derivation of self-consistent equations \cite{2008_Cherroret}. The agreement between theory and numerical simulations is good in the presence of absorption as well, which has a particular importance in the context of the recent quest for Anderson localization of classical waves that heavily relies on confrontation of experimental results with the self-consistent theory \cite{2004_Skipetrov,2008_van_Tiggelen_Nature,2006_Maret_PRL, 2006_Skipetrov_dynamics,2003_Genack}. Deep in the localized regime ($g_0 < 0.5$), the self-consistent theory loses its quantitative accuracy, but still yields qualitatively correct results (exponential decay of conductance with the length of the waveguide and of the diffusion coefficient $D$ with the distance from waveguide boundaries). It would be extremely interesting to see if the ability of the self-consistent theory to provide quantitative predictions still holds in three-dimensional systems where a mobility edge exists. In particular, the immediate proximity of the mobility edge is of special interest.

\textit{Note added.}
After this paper was submitted for publication, a related preprint appeared \cite{2010_Tian}. In particular, the authors of that work show that the self-consistent theory does not apply to 1D disordered media, which is consistent with our results because $g_0 \sim \ell/L$ is always small in 1D, provided that the condition $L \gg \ell$ assumed in this paper is fulfilled.

\acknowledgments
We thank Bart van Tiggelen for useful comments. The work at Missouri S\&T was supported by the National Science Foundation Grant No. DMR-0704981. The numerical results obtained at the Tera-Grid, award Nos. DMR-090132 and DMR-100030. S.E.S. acknowledges financial support of the French ANR (Project No. 06-BLAN-0096 CAROL).




\begin{thebibliography}{99}

\bibitem{1958_Anderson}
P.~W.~Anderson, Phys. Rev. \textbf{109}, 1492 (1958)

\bibitem{1984_John_prl}
S.~John, Phys.  Rev. Lett. \textbf{53}, 2169  (1984);
P.~W.~Anderson,  Philos. Mag. B \textbf{52}, 505 (1985)

\bibitem{billy08}
J. Billy {\em et al.},
Nature (London) \textbf{453}, 891 (2008);
G. Roati {\em et al.},
Nature (London) \textbf{453}, 895 (2008).

\bibitem{2007_Akkermans_book}
E.~Akkermans and  G.~Montambaux,  \emph{Mesoscopic Physics of Electrons and Photons}  (Cambridge University Press, 2007)

\bibitem{1991_Genack}
A.~Z.~Genack and  N.~Garcia,  Phys. Rev. Lett. \textbf{66}, 2064 (1991);
R.~Weaver,  Phys. Rev. B \textbf{47}, 1077 (1993);
D.~S.~Wiersma,  P.~Bartolini,  A.~Lagendijk,  and R.~Righini,  Nature \textbf{390},671 (1997);
F.~Scheffold,  R.~Lenke,  R.~Tweer, and  G.~Maret,  Nature \textbf{398},206 (1999)

\bibitem{2000_chabanov_nature}
A.~A.~Chabanov,  M.~Stoytchev,  and A.~Z. Genack, Nature  \textbf{404}, 850  (2000)

\bibitem{2003_Genack}
A.~A.~Chabanov,  Z.~Q.~Zhang, and  A.~Z.~Genack,  Phys. Rev. Lett. \textbf{90},  203903 (2003);
Z.~Q.~Zhang,  A.~A.~Chabanov,  S.~K.~Cheung,  C.~H.~Wong, and  A.~Z.~Genack,  Phys. Rev. B \textbf{79},  144203 (2009)

\bibitem{2006_Maret_PRL}
M.~St\"{o}rzer,  P.~Gross,  C.~Aegerter, and  G.~Maret,  Phys. Rev. Lett.\textbf{96},  063904 (2006);
T.~Schwartz,  G.~Bartal,  S.~Fishman, and  M.~Segev,  Nature \textbf{446}, 52  (2007)

\bibitem{2008_van_Tiggelen_Nature}
H.~Hu,  A.~Strybulevych,  J.~H.~Page,  S.~E.~Skipetrov,  and B.~A.~van  Tiggelen, Nat. Phys. \textbf{4}, 945 (2008)

\bibitem{schwartz07}
T. Schwartz {\em et al.},
Nature (London) \textbf{446}, 52 (2007);
Y. Lahini {\em et al.},
Phys. Rev. Lett. \textbf{100}, 013906 (2008).

\bibitem{2000_van_Tiggelen}
B.~A.~van Tiggelen,  A.~Lagendijk,  and D.~S.  Wiersma, Phys. Rev. Lett.  \textbf{84}, 4333  (2000)

\bibitem{2004_Skipetrov}
S.~E.~Skipetrov  and B.~A.~van  Tiggelen, Phys. Rev. Lett.  \textbf{92}, 113901  (2004)

\bibitem{2006_Skipetrov_dynamics}
S.~E.~Skipetrov  and B.~A.~van  Tiggelen, Phys. Rev. Lett.  \textbf{96}, 043902  (2006)

\bibitem{2008_Cherroret}
N.~Cherroret and  S.~E.~Skipetrov,  Phys. Rev. E \textbf{77},  046608 (2008)

\bibitem{1980_Vollhardt_Wolfle}
D.~Vollhardt and  P.~W\"olfle,  Phys. Rev. B \textbf{22}, 4666 (1980);
J.~Kroha,  C.~M.~Soukoulis,  and P.~W\"olfle,  Phys. Rev. B \textbf{47}, 11093 (1993)

\bibitem{2009_Cherroret}
N.~Cherroret,  S.~E.~Skipetrov, and B.~A.~van  Tiggelen, Phys. Rev. B \textbf{80},  037101 (2009);
N.~Cherroret,  S.~E.~Skipetrov, and B.~A.~van  Tiggelen (2008), arXiv:0810.0767

\bibitem{2008_Tian}
C. Tian,
Phys. Rev. B \textbf{77}, 064205 (2008)

\bibitem{2000_Mirlin}
A.~Mirlin,  Phys. Rep. \textbf{326}, 259 (2000)

\bibitem{1999_van_Rossum}
M.~C.~van Rossum and T.~M. Nieuwenhuizen, Rev. Mod. Phys.  \textbf{71}, 313  (1999)

\bibitem{1997_Beenakker}
C.~W.~Beenakker,  Rev. Mod. Phys. \textbf{69},731 (1997)

\bibitem{2007_Froufe-Perez_PRE}
L.~S.~Froufe-P\'{e}rez,  M.~Y\'{e}pez,  P.~A.~Mello, and  J.~J.~S\'{a}enz,  Phys. Rev. E \textbf{75},  031113 (2007)

\bibitem{2010_Payne_closed}
B.~Payne,  T.~Mahler, and  A.~Yamilov  (2010), unpublished

\bibitem{1999_yamilov_selfembed}
L.~I.~Deych,  A.~Yamilov, and  A.~A.~Lisyansky,  Phys. Rev. B \textbf{59}, 11339 (1999)

\bibitem{2006_Yamilov_conductance}
A.~Yamilov and  H.~Cao,  Phys. Rev. E \textbf{74},  056609 (2006)

\bibitem{1953_Morse}
P.~M.~Morse and  H.~Feshbach,  \emph{Methods of Theoretical Physics}  (McGraw-Hill, New York,  1953)

\bibitem{2010_Tian}
C.S. Tian, S.K. Cheung and Z.Q. Zhang,
arXiv:1005.0951

\end{thebibliography}
\end{document}